\documentclass[11pt]{article}
\usepackage[a4paper]{geometry}
\usepackage{amssymb,amsmath,amsthm,graphicx}
\usepackage[numbers]{natbib}

\DeclareMathOperator*{\med}{median}
\newcommand{\R}{\mathbb{R}}
\newcommand{\bzero}{\boldsymbol 0}
\newcommand{\ba}{\boldsymbol a}

\newcommand{\bu}{\boldsymbol u}

\newcommand{\bx}{\boldsymbol x}

\newcommand{\bet}{\boldsymbol \eta}

\newcommand{\btheta}{\boldsymbol \theta}
\newcommand{\bxi}{\boldsymbol \xi}

\begin{document}
\large

\title{Statistical depth meets computational geometry:\\ 
       a short survey}
\author{Peter J. Rousseeuw and Mia Hubert}
\date{August 15, 2015}

\maketitle

\begin{abstract}
During the past two decades there has been a lot of 
interest in developing statistical depth notions that
generalize the univariate concept of ranking to
multivariate data. The notion of depth has also been
extended to regression models and functional data.
However, computing such depth functions as well as
their contours and deepest points is not trivial.
Techniques of computational geometry appear to be 
well-suited for the development of such algorithms.
Both the statistical and the computational geometry 
communities have done much work in this direction,
often in close collaboration.
We give a short review of this work, focusing mainly
on depth and multivariate medians, and end by listing
some other areas of statistics where computational geometry 
has been of great help in constructing efficient algorithms.\\
\end{abstract}

\section{Depth notions}
\noindent
A data set consisting of $n$ univariate points is usually 
ranked in ascending or descending order. Univariate order 
statistics (i.e., the `$k$th smallest value out of $n$') 
and derived quantities have been studied extensively.
The median is defined as the order statistic of rank 
$(n+1)/2$ when $n$ is odd, and as the average of the order 
statistics of ranks $n/2$ and $(n+2)/2$ when $n$ is even. 
The median and any other order statistic of a univariate 
data set can be computed in $O(n)$ time.
Generalization to higher dimensions is, however, not 
straightforward.

Alternatively, univariate points may be ranked from the 
outside inward by assigning the most extreme data points 
depth 1, the second smallest and second largest data points 
depth 2, etc.
The deepest point then equals the usual median of the sample.
The advantage of this type of ranking is that it can be 
extended to higher dimensions more easily. This section 
gives an overview of several possible generalizations of 
depth and the median to multivariate settings. 
Surveys of statistical applications of multivariate data 
depth may be found in~\cite{LPS99}, \cite{ZS00}, 
and~\cite{Mos13}.

\subsection{Halfspace location depth}
\begin{figure}[!ht]
\centering
\includegraphics[width=0.80\textwidth]
                {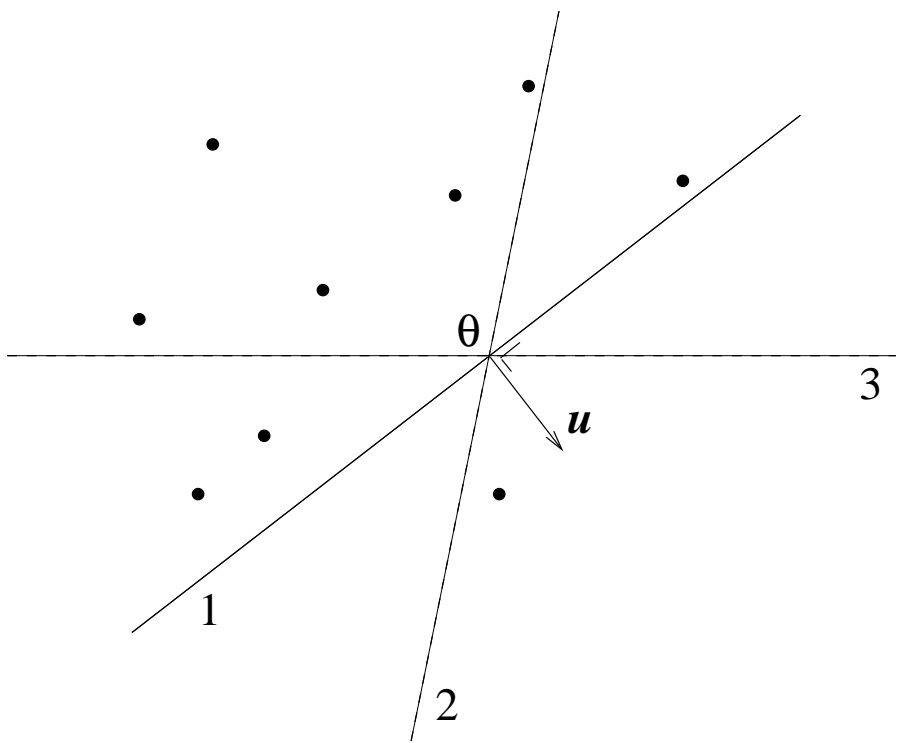}
\caption{Illustration of the bivariate halfspace depth. 
Here $\btheta$ (which is not a data point itself) has depth 1 
because the halfspace determined by $\bu$ contains only one 
data point.}
\label{fig:ldep}
\end{figure}

\noindent
Let $X_n=\{\bx_1,\ldots,\bx_n\}$ be a finite set of data 
points in $\R^d$.
The {\em Tukey depth\/} or {\em halfspace depth\/}
(introduced by~\cite{Tuk75} and further developed 
by~\cite{DG92}) of any point $\btheta$ in $\R^d$ (not 
necessarily a data point) determines how central the 
point is inside the data cloud.
The halfspace depth of $\btheta$ is defined as the minimal 
number of data points in any closed halfspace determined 
by a hyperplane through $\btheta$:
\[hdepth(\btheta; X_n) = \min_{\|\bu\|=1}\
\#\{i; \bu^\tau \bx_i \geqslant \bu^\tau \btheta\}.\]
Thus, a point lying outside the convex hull of $X_n$ has 
depth $0$, and any data point has depth at least 1.
Figure~\ref{fig:ldep} illustrates this definition for $d=2$.

The set of all points 
with depth $\geqslant k$ is called the $k$th depth region $D_k$.
The halfspace depth regions form a sequence of nested 
polyhedra. Each $D_k$ is the intersection of all 
halfspaces containing at least $n-k+1$ data points.
Moreover, every data point must be a vertex of one or more 
depth regions. The point with maximal halfspace depth is called 
the {\em Tukey median}. When the innermost depth region is 
larger than a singleton, the Tukey median is defined as 
its centroid. This makes the Tukey median unique by 
construction.

\begin{figure}[!ht]
\centering
\includegraphics[width=0.80\textwidth]
                {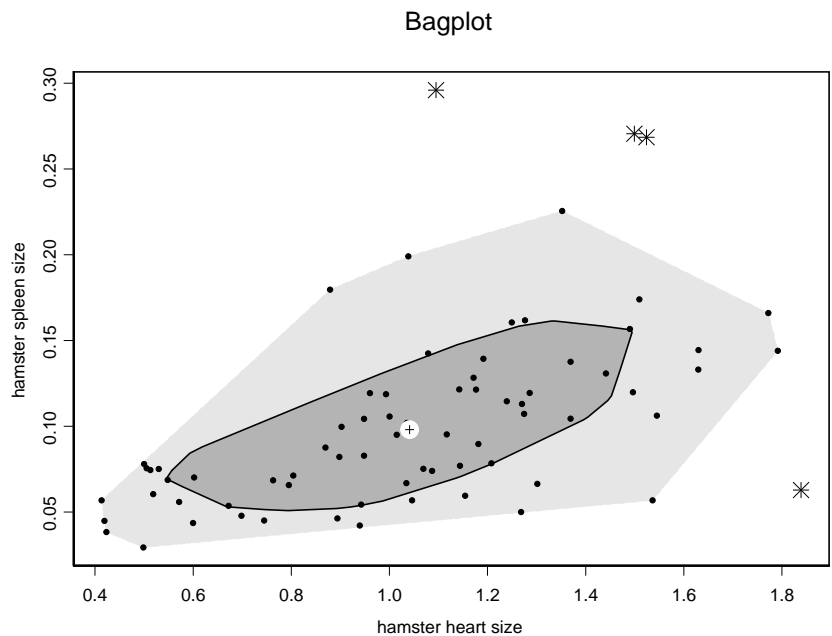}
\caption{Bagplot of the heart and spleen size of 73 hamsters.}
\label{fig:bag}
\end{figure}

Note that the depth regions give an indication of the 
shape of the data cloud. Based on this idea one can 
construct the {\em bagplot\/} \cite{RRT99}, a bivariate 
version of the univariate boxplot. Figure~\ref{fig:bag}
shows such a bagplot. The cross in the white disk is the 
Tukey median. The dark area is an interpolation between two
subsequent depth regions, and contains 50\% of the data. 
This area (the ``bag'') gives an idea of the shape of the 
majority of the data cloud.
Inflating the bag by a factor of 3 relative to the Tukey 
median yields the ``fence'' (not shown), and data points 
outside the fence are called outliers and marked by stars. 
Finally, the light gray area is the convex hull of the
non-outlying data points.

More generally, in the multivariate case one can define
the {\em bagdistance\/} \cite{HRS15b} of a point $\bx$
relative to the Tukey median and the bag. Assume that 
the Tukey median lies in the interior of the bag, not on
its boundary (this excludes degenerate cases). 
Then the bagdistance is the smallest real number $\lambda$ 
such that the bag inflated (or deflated) by $\lambda$ 
around the Tukey median contains the point $\bx$. 
When the Tukey median equals $\bzero$, it is shown 
in~\cite{HRS15b} that the bagdistance satisfies all 
axioms of a norm except that $||a \bx|| = |a | ||\bx||$
only needs to hold when $a \geqslant 0$.
The bagdistance is used for outlier 
detection~\cite{HRS15a} and statistical
classification~\cite{HRS15b}. 

An often used criterion to judge the robustness of an 
estimator is its {\em breakdown value}. The breakdown value 
is the smallest fraction of data points that we need to 
replace in order to move the estimator of the contaminated 
data set arbitrarily far away.
The classical mean of a data set has breakdown value zero 
since we can move it anywhere by moving one observation. 
Note that for any estimator which is equivariant 
for translation (which is required to call it a location 
estimator) the breakdown value can be at most 1/2. (If we 
replace half of the points by a far-away translation image 
of the remaining half, the estimator cannot distinguish 
which were the original data.)

The Tukey depth and the corresponding median have good 
statistical properties. The Tukey median $T^*$ is a 
location estimator with breakdown
value $\varepsilon_n(T^*;X_n)$ $\geqslant 1/(d+1)$ for any 
data set in general position. This means that it remains in a 
predetermined bounded region unless $n/(d+1)$ or more data 
points are moved. At an elliptically symmetric distribution 
the breakdown value becomes 1/3 for large $n$, irrespective 
of $d$. Moreover, the halfspace depth is invariant under all 
nonsingular affine transformations of the data, making the 
Tukey median affine equivariant. Since data transformations 
such as rotation and rescaling are very common in 
statistics, this is an important property.
The statistical asymptotics of the Tukey median have been 
studied in~\cite{BH99}.

The need for fast algorithms for the halfspace depth has 
only grown over the years, since it is currently being 
applied to a variety of settings such as nonparametric 
classification~\cite{LCL12}.
A related development is the fast growing field of functional
data analysis, where the data are functions on a univariate
interval (e.g.\ time or wavelength) or on a rectangle
(e.g.\ surfaces, images). 
Often the function values are themselves multivariate.
One can then define the depth of a curve (surface) by
integrating the depth over all points as in~\cite{Cla14}.
This functional depth can again be used for outlier 
detection and classification~\cite{HRS15a, HRS15b}, but it 
requires computing
depths in many multivariate data sets instead of just one.

{\bf Remark: centerpoints.} 
There is a close relationship between the Tukey 
depth and centerpoints, which have been long studied in 
computational geometry. In fact, Tukey depth extends the 
notion of centerpoint. A {\em centerpoint\/} is any point 
with halfspace depth $\geqslant \lceil n/(d+1)\rceil$.
A consequence of Helly's theorem is 
that there always exists at least one centerpoint, so the 
depth of the Tukey median cannot be less than
$\lceil n/(d+1)\rceil$.

\subsection{Other location depth notions}

\begin{enumerate}
\item Simplicial depth (\cite{Liu90}).
The depth of $\btheta$ equals the number of simplices formed 
by $d+1$ data points that contain $\btheta$. Formally,
\[sdepth(\btheta; X_n) = \#\{(i_1,\ldots,i_{d+1});
\btheta \in S[\bx_{i_1},\ldots,\bx_{i_{d+1}}]\}.\]
The simplicial median is affine equivariant with a breakdown 
value bounded above by $1/(d+2)$. Unlike halfspace depth, its
depth regions need not be convex.
\item Oja depth (\cite{Oja83}).
This is also called simplicial volume depth:
\[odepth(\btheta; X_n) = \bigl(1+\sum_{(i_1,\ldots,i_{d})}\
\{volume\ S[\btheta, \bx_{i_1},\ldots,\bx_{i_{d}}]\} \bigr)^{-1}.\]
The corresponding median is also affine equivariant, but has
zero breakdown value.
\item Projection depth.
We first define the {\em outlyingness\/} (\cite{DG92})
of any point $\btheta$ relative to the data set $X_n$ as
\[ O(\btheta;X_n)=\max_{\|\bu\|=1}
\frac{|\bu^\tau \btheta - \med_i\{\bu^\tau \bx_i\}|}
{\mbox{MAD}_i\{\bu^\tau \bx_i \} }, \]
where the median absolute deviation (MAD) of a univariate 
data set\linebreak
$\{y_1, \ldots, y_n\}$ is the statistic
$\mbox{MAD}_i\{y_i\}=\med_i|y_i-\med_j\{y_j\}|$.
The outlyingness is small for centrally located points and
increases if we move toward the boundary of the data cloud.
Instead of the median and the MAD, also another pair
$(T,S)$ of a location and scatter estimate may be chosen.
This leads to different notions of projection depth, all 
defined as \[pdepth(\btheta; X_n) = (1+O(\btheta;X_n))^{-1}.\]
General projection depth is studied in~\cite{Zuo03}.
When using the median and the MAD, the projection depth 
has breakdown value 1/2 and is affine equivariant.
Its depth regions are convex.
\item Spatial depth (\cite{Ser02}). 
Spatial depth is related to multivariate quantiles proposed 
in \cite{Cha96}:
\[spdepth(\btheta; X_n) = 1-
  \left\|\frac{1}{n}\sum_{i=1}^n \frac{\bx_i-\btheta}{\|\bx_i-\btheta \|} \right\|  \]
The spatial median is also called the $L^1$ median (\cite{Gow74}).
It has breakdown value 1/2, but is not affine equivariant 
(it is only equivariant with respect to translations, 
multiplication by a scalar factor, and orthogonal 
transformations). For a recent survey on the computation
of the spatial median see~\cite{FFC12}.
\end{enumerate}

A comparison of the main properties of the different 
location depth medians is given in Table~\ref{tab:prop}.

\begin{table}[!ht]
\begin{center}
\baselineskip=11pt
\renewcommand{\arraystretch}{.9}
\caption{Comparison of several location depth medians}
\vskip3mm
\label{tab:prop}
{\begin{tabular}{| l | c | c | }
	\hline
\rule[-4pt]{0pt}{13pt}{MEDIAN}
	& {BREAKDOWN VALUE}
	& {AFFINE EQUIVARIANCE}
	\\ \hline
\rule[0pt]{0pt}{9pt}Tukey
	& worst-case $1/(d+1)$
	& yes
	\\
	& typically $1/3$
	& \\
Simplicial
	& $\leqslant 1/(d+2)$
	& yes
	\\
Oja
	& $2/n \approx 0$
	& yes
	\\
Projection
	& $1/2$
	& yes
	\\
Spatial
	& $1/2$
	& no
	\\ \hline
\end{tabular}}
\renewcommand{\arraystretch}{1}
\baselineskip=12pt
\end{center}
\end{table}

\subsection{Arrangement and regression depth}

\begin{figure}[!ht]
\centering
\includegraphics[angle=-90,width=1.0\textwidth]
                {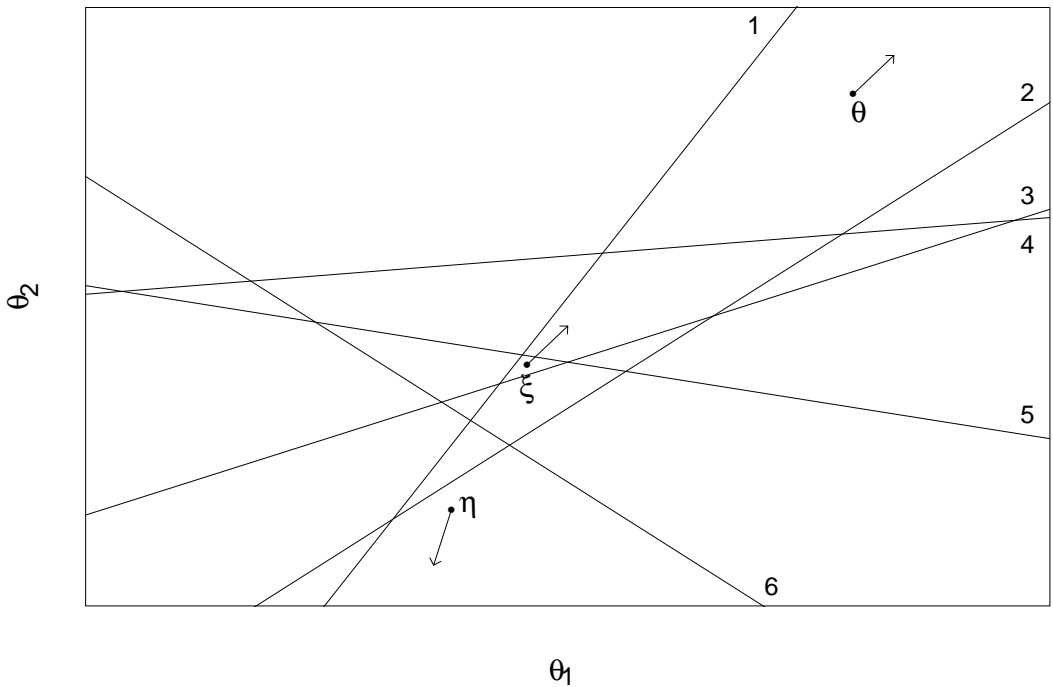}
\caption{Example of arrangement depth. In this arrangement 
of lines, the points $\btheta$ and $\bet$ have arrangement 
depth 0, whereas $\bxi$ has arrangement depth 2. (See 
Figure~\ref{fig:rdepd2} for the dual plot.)}
\label{fig:rdepd1}
\end{figure}

\noindent
Following~\cite{RH99b} we now define the depth of a 
point relative to an arrangement of hyperplanes.
A point $\btheta$ is said to have zero arrangement depth
if there exists a 
ray $\{\btheta+\lambda \bu; \lambda\geqslant 0\}$ 
that does not cross any of the hyperplanes $h_i$ in 
the arrangement. (A hyperplane parallel to the ray is 
counted as intersecting at infinity.) The arrangement 
depth of any point $\btheta$ is defined as the minimum 
number of hyperplanes intersected by any ray from $\btheta$.
Figure~\ref{fig:rdepd1} shows an arrangement of lines.
In this plot, the points $\btheta$ and $\bet$ have 
arrangement depth 0 and the point $\bxi$ has arrangement
depth 2. The arrangement depth is always constant on 
open cells and on cell edges. It was shown (\cite{RH99b}) 
that any arrangement of lines in the plane encloses a 
point with arrangement depth at least $\lceil n/3 \rceil$, 
giving rise to a new type of ``centerpoints.''

\begin{figure}[!ht]
\centering
\includegraphics[angle=-90,width=1.0\textwidth]
                {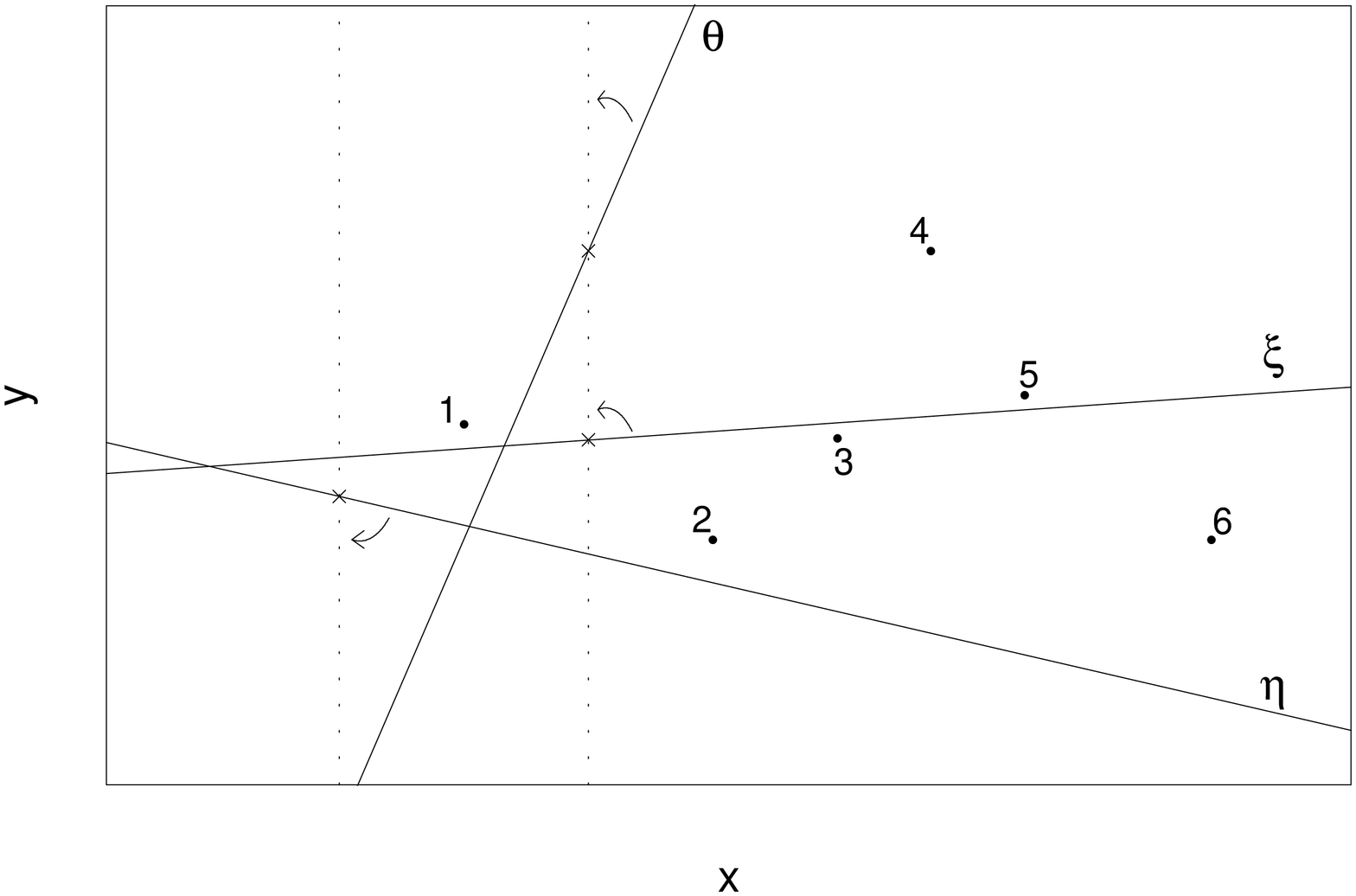}
\caption{Example of the regression depth of a line in a bivariate
configuration of points. The lines $\btheta$ and $\bet$ have 
regression depth 0, whereas the line $\bxi$ has regression 
depth 2. (This is the dual of Figure~\ref{fig:rdepd1}.)}
\label{fig:rdepd2}
\end{figure}

This notion of depth was originally defined (\cite{RH99}) 
in the dual, as the depth of a regression hyperplane 
$H_{\btheta}$ relative to a point configuration
of the form $Z_n=\{(\bx_1,y_1),\ldots,(\bx_n,y_n)\}$ 
in $\R^{d+1}$. 
Regression depth ranks hyperplanes according to how well 
they fit the data in a regression model, with $\bx$ 
containing the predictor variables and $y$ the response. 
A vertical hyperplane (given by
$\ba^\tau \bx=$ constant), which cannot be used to 
predict future response values, is called a ``nonfit'' 
and assigned regression depth 0.
The regression depth of a general hyperplane $H_{\btheta}$ 
is found by rotating $H_{\btheta}$ in a continuous movement
until it becomes vertical.
The minimum number of data points that is passed in 
such a rotation is called the regression depth 
of $H_{\btheta}$.
Figure~\ref{fig:rdepd2} is the dual representation of
Figure~\ref{fig:rdepd1}.
(For instance, the line $\btheta$ has slope $\theta_1$ 
and intercept $\theta_2$ and corresponds to the point
$(\theta_1,\theta_2)$ in Figure~\ref{fig:rdepd1}.)
The lines $\btheta$ and $\bet$ have regression depth 0, 
whereas the line $\bxi$ has regression depth 2.

In statistics one is interested in the 
{\em deepest fit\/} or regression depth median, because 
this is a line (hyperplane) about which the data are 
well-balanced. The statistical properties of regression 
depth and the deepest fit are very similar to those of 
the Tukey depth and median. The bounds on the maximal
depth are almost the same. Moreover, for both depth 
notions the value of the maximal depth can be used to 
characterize the symmetry of the 
distribution (\cite{RS04}).
The breakdown value of the deepest fit is at least 
$1/(d+1)$ and under linearity of the conditional median 
of $y$ given $\bx$ it converges to 1/3. In the next 
section, we will see that the optimal complexities for 
computing the depth and the median are also comparable
to those for halfspace depth. 
For a detailed comparison of the properties of halfspace 
and regression depth, see~\cite{HRV01}.

The arrangement depth region $D_k$ is defined in the 
primal, as the set of points with arrangement depth at 
least $k$. Contrary to the Tukey depth, these depth 
regions need not be convex. But nevertheless it was 
proved that there always exists a point with arrangement 
depth at least $\lceil n/(d+1)\rceil$ (\cite{ABEp00}).
An analysis-based proof was given in~\cite{Miz02}.

{\bf Remark: arrangement levels.} 
Arrangement depth is undirected (isotropic) in the 
sense that it is defined as a minimum over all possible 
directions. If we restrict ourselves to vertical 
directions $\bu$ (i.e., up or down), we obtain the usual 
levels of the arrangement known in combinatorial
geometry. 
The absence of preferential directions makes arrangement 
depth invariant under affine transformations.

\section{Computing depth}
\noindent Although the definitions of depth are 
intuitive, the computational aspects can be quite 
challenging. The calculation of depth regions and 
medians is computationally intensive, especially for 
large data sets in higher dimensions. In statistical 
practice such data are quite common, and therefore 
reliable and efficient algorithms are needed. 
For the bivariate case several algorithms have been 
developed early on. The computational aspects of depth 
in higher dimensions are currently being explored.

Algorithms for depth-related measures are often more 
complex for data sets which are not in general position 
than for data sets in general position. For example,
the boundaries of subsequent halfspace depth regions are 
always disjoint when the data are in general position, 
but this does not hold for nongeneral position.
Preferably, algorithms should be able to handle both the 
general position case and the nongeneral position case 
directly. As a quick fix, algorithms which were made for 
general position can also be applied in the other case 
if one first adds small random errors to the data 
points. For large data sets, this `dithering' will have 
a limited effect on the results.

\subsection{Bivariate algorithms}
\noindent
Table~\ref{tab:ldep} gives an overview of algorithms,
each of which has been implemented, to compute the
depth in a given point $\btheta$ in $\R^2$.
These algorithms are time-optimal, since the problem of 
computing these bivariate depths has an $\Omega(n\log n)$ 
lower bound (\cite{ACGp01}, \cite{LS00b}).

The algorithms for halfspace and simplicial depth are 
based on the same technique. First, data points are 
radially sorted around $\btheta$. Then a line through $\btheta$
is rotated. The depth is calculated by counting the
number of points that are passed by the rotating line 
in a specific manner. 
The planar arrangement depth algorithm is easiest to 
visualize in the regression setting. To compute the 
depth of a hyperplane $H_{\btheta}$ with coefficients $\btheta$,
the data are first sorted along the $x$-axis. A vertical 
line $L$ is then moved from left to right and each time 
a data point is passed, the number of points above and 
below $H_{\btheta}$ on both sides of $L$ is updated.

\begin{table}[!ht]
\begin{center}
\baselineskip=11pt
\renewcommand{\arraystretch}{.9}
\caption{Computing the depth of a bivariate point.}
\label{tab:ldep}
\vskip3mm
{\begin{tabular}{| l | c | c | }
	\hline
\rule[-4pt]{0pt}{13pt}{DEPTH}
	& {TIME COMPLEXITY}
	& {SOURCE}
	\\ \hline
\rule[0pt]{0pt}{9pt}Tukey depth
	& $O(n \log n)$
	& \cite{RR96}
	\\
Simplicial depth
	& $O(n \log n)$
	& \cite{RR96}
	\\ 
Arrangement/regression depth
	& $O(n \log n)$
	& \cite{RH99}
	\\
\hline
\end{tabular}}
\renewcommand{\arraystretch}{1}
\baselineskip=12pt
\end{center}
\end{table}

In general, computing a median is harder than 
computing the depth in a point, because typically there 
are many candidate points. For instance, for the 
bivariate simplicial median the currently best algorithm 
requires $O(n^4)$ time, whereas its corresponding depth 
needs only $O(n \log n)$. The simplicial median seems 
difficult to compute because there are $O(n^4)$ candidate 
points (namely, all intersections of lines passing through 
two data points) and the simplicial depth regions have 
irregular shapes, but of course a faster algorithm may yet 
be found.

Fortunately, in several important cases the median can be 
computed without computing the depth of individual points.
A linear-time algorithm to compute a bivariate 
centerpoint was described in~\cite{JM94}.
Table~\ref{tab:lmed} gives an overview of algorithms to 
compute bivariate depth-based medians.
For the bivariate Tukey median the lower bound 
$\Omega(n\log n)$ was proved in \cite{LS00}, and the 
currently fastest algorithm takes $O(n \log^3 n)$ time
(\cite{LS03}).
The lower bound $\Omega(n\log n)$ also holds for the
median of arrangement (regression) depth as shown 
by~\cite{LS03b}. 
Fast algorithms were devised by~\cite{LS03b} 
and~\cite{VMRp08}.

\begin{table}[!ht]
\begin{center}
\baselineskip=11pt
\renewcommand{\arraystretch}{.9}
\caption{Computing the bivariate median.}
\label{tab:lmed}
\vskip3mm
{\begin{tabular}{| l | c | c | }
	\hline
\rule[-4pt]{0pt}{13pt}{MEDIAN}
	& {TIME COMPLEXITY}
	& {SOURCE}
	\\ \hline
\rule[0pt]{0pt}{9pt}Tukey median
	& $O(n \log^3 n)$
	& \cite{LS03}
	\\
Simplicial median
	& $O(n^4)$
	& \cite{ALSp01}
	\\
Oja median
	& $O(n \log^3 n)$
	& \cite{ALSp01}
	\\ 
Regression depth median
	& $O(n \log n)$
	& \cite{LS03b}
	\\
\hline
\end{tabular}}
\renewcommand{\arraystretch}{1}
\baselineskip=12pt
\end{center}
\end{table}

The computation of bivariate halfspace depth regions 
has also been studied.
The first algorithm~\cite{RR96b} required $O(n^2 \log n)$
time per depth region. 
An algorithm to compute all regions in $O(n^2)$ time is 
constructed and implemented in~\cite{MRRp03}. This 
algorithm thus also yields the Tukey median. It is based 
on the dual arrangement of lines where topological sweep 
is applied. A completely different approach is
implemented in~\cite{KMV02}. They make direct use of the 
graphics hardware to approximate the depth regions of a 
set of points in $O(nW+W^3)+nCW^2/512$ time, where the 
pixel grid is of dimension $(2W+1)\times(2W+1)$.
Recently, \cite{BRS11} constructed an algorithm to
update halfspace depth and its regions when points
are added to the data set.

\subsection{Algorithms in higher dimensions}
\noindent
The first algorithms to compute the halfspace and 
regression depth of a given point in $\R^d$ with $d>2$ 
were constructed in~\cite{RS98} and 
require $O(n^{d-1}\log n)$ time. 
The main idea was to use projections onto a lower-dimensional 
space. This reduces the problem to computing bivariate 
depths, for which the existing algorithms have optimal
time complexity.
In~\cite{BCI+08} theoretical output-sensitive algorithms for 
the halfspace depth are proposed.
An interesting computational connection between halfspace
depth and multivariate quantiles was provided 
in~\cite{HPS10} and~\cite{KM12}.
More recently, \cite{DM14} provided a generalized version of 
the algorithm of~\cite{RS98} together with C++ code.
For the depth regions of halfspace depth
in higher dimensions an algorithm was recently 
proposed in~\cite{LMM14}.

For the computation of projection depth see~\cite{LZ14}.
The simplicial depth of a point in $\R^3$ can be computed 
in $O(n^2)$ time, and in $\R^4$ the fastest algorithm 
needs $O(n^4)$ time~\cite{CO01}.
For higher dimensions, no better algorithm is known than 
the straightforward $O(n^{d+1})$ method to compute all 
simplices.

When the number of data points and dimensions are such
that the above algorithms become infeasible, one can
resort to approximate algorithms.
For halfspace depth such approximate algorithms were 
proposed in~\cite{RS98} and~\cite{CMW13}.
An approximation to the Tukey median using steepest descent
can be found in~\cite{SR00}.
In~\cite{VRHp02} an algorithm is described to approximate 
the deepest regression fit in any dimension.

\section{Some other statistical techniques benefitting 
         from computational geometry}
\noindent
Computational geometry has provided fast and reliable 
algorithms for many other statistical techniques.

Linear regression is a frequently used statistical technique. 
The ordinary least squares regression, minimizing the sum of 
squares of the residuals, is easy to calculate, but produces
unreliable results whenever one or more outliers are present 
in the data. 
Robust alternatives are often computationally intensive. 
We here give some examples of regression methods for which 
geometric or combinatorial algorithms have been constructed.
\begin{enumerate}
\item $L^1$ regression.
This well-known alternative to least squares regression 
minimizes the sum of the absolute values of the residuals, 
and is robust to vertical outliers. Algorithms for $L^1$ 
regression may be found in, e.g.,~\cite{YKIp88} and~\cite{PK97}.
\item Least median of squares (LMS) regression (\cite{Rou84}). 
This method minimizes the median of the squared residuals and 
has a breakdown value of 1/2.
To compute the bivariate LMS line, an $O(n^2)$ algorithm using topological sweep has been developed~\cite{ES90}.
An approximation algorithm for the LMS line was constructed
in~\cite{MNRp97}.
The recent algorithm of~\cite{BM14} uses mixed integer 
optimization.
\item Median slope regression (\cite{The50}, \cite{Sen68}).
This bivariate regression technique estimates the slope
as the median of the slopes of all lines through two data 
points.
An algorithm with optimal complexity $O(n \log n)$ is 
given in~\cite{BC98}, and a more practical randomized 
algorithm in~\cite{DMN92}.
\item Repeated median regression (\cite{Sie82}).
Median slope regression takes the median over all couples
($d$-tuples in general) of data points. Here, this median is 
replaced by $d$ nested medians. For the bivariate repeated 
median regression line, \cite{MMN98} provide an efficient 
randomized algorithm.
\end{enumerate}

The aim of cluster analysis
is to divide a data set into clusters of similar objects.
Partitioning methods divide the data into $k$ groups.
Hierarchical methods construct a complete clustering tree, 
such that each cut of the tree gives a partition of the 
data set.
A selection of clustering methods with accompanying 
algorithms is presented in~\cite{SHR97}.
The general problem of partitioning a data set into groups 
such that the partition minimizes a given error 
function $f$ is NP-hard.
However, for some special cases efficient algorithms exist.
For a small number of clusters in low dimensions, exact 
algorithms for partitioning methods can be constructed. 
Constructing clustering trees is also closely related to 
geometric problems (see e.g., \cite{Epp97}, \cite{Epp98}).

\section{Other surveys}
\noindent All results not given an explicit reference 
above may be traced in these surveys.
\begin{trivlist}
\item []
\cite{Mos13}: A survey of multivariate data depth and its 
statistical applications.
\item []
\cite{Sha76}: An overview of the computational complexities 
of basic statistics problems like ranking, regression, and 
classification.
\item []
\cite{Sma90}: An overview of several multivariate medians 
and their basic properties.
\item []
\cite{ZS00}: A classification of multivariate data depths 
based on their statistical properties.
\end{trivlist}


\begin{thebibliography}{}

\bibitem[ACG{\etalchar{+}}02]{ACGp01} 
G.~Aloupis, C.~Cort\'{e}s, F.~G\'omez, M.~Soss, and G.T.~Toussaint.%
\newblock Lower bounds for computing statistical depth.
\newblock {\em Comput. Stat. Data Anal.}, 40:223--229, 2002. 

\bibitem[ALS{\etalchar{+}}03]{ALSp01} 
G.~Aloupis, S.~Langerman, M.~Soss, and G.T.~Toussaint.%
\newblock Algorithms for bivariate medians and a Fermat-Torricelli problem 
for lines.
\newblock {\em Comput. Geom. Theory Appl.}, 26:69--79, 2003. 

\bibitem[ABE{\etalchar{+}}00]{ABEp00} 
N.~Amenta, M.~Bern, D.~Eppstein, and S.-H.~Teng. %
\newblock Regression depth and center points. 
\newblock {\em Discrete Comput. Geom.}, 23:305--323, 2000.

\bibitem[BH99]{BH99} 
Z.-D.~Bai and X.~He. %
\newblock Asymptotic distributions of the maximal depth 
estimators for regression and multivariate location.
\newblock {\em Ann. Statist.}, 27:1616--1637, 1999.

\bibitem[BM14]{BM14}
D.~Bertsimas and R.~Mazumder.
\newblock Least quantile regression via modern optimization.
\newblock {\em Ann. Statist.}, 42:2494--2525, 2014.

\bibitem[BCI+08]{BCI+08}
D.~Bremner, D.~Chen, J.~Iacono, S.~Langerman, and P.~Morin.
\newblock Output-sensitive algorithms for {T}ukey depth
          and related problems.
\newblock {\em Stat. Comput.}, 18:259--266, 2008.

\bibitem[BC98]{BC98} 
H.~Br\"{o}nnimann and B.~Chazelle. %
\newblock Optimal slope selection via cuttings.
\newblock {\em Comput. Geom.}, 10:23--29, 1998.

\bibitem[BRS11]{BRS11}
M.~Burr, E.~Rafalin, and S.L.~Souvaine.
\newblock Dynamic maintenance of halfspace depth for points
          and contours.
\newblock arXiv:1109.1517 (2011).

\bibitem[Cha96]{Cha96}
P.~Chaudhuri.
\newblock On a geometric notion of quantiles for multivariate data.
\newblock {\em J. Amer. Statist. Assoc.},  91:862--872, 1996.

\bibitem[CMW13]{CMW13}
D.~Chen, P.~Morin, and U.~Wagner.
\newblock Absolute approximation of Tukey depth: theory
          and experiments.
\newblock {\em Comput. Geom.}, 46:566--573, 2013. 

\bibitem[CO01]{CO01} 
A.Y.~Cheng and M.~Ouyang. %
\newblock On algorithms for simplicial depth.
\newblock In {\em Proc. 13th Canadian Conf. on Comp. Geom.},
          pages 53--56, Waterloo, 2001.

\bibitem[Cla14]{Cla14}
G. Claeskens, M. Hubert, L. Slaets, and K. Vakili.
\newblock Multivariate functional halfspace depth.
\newblock {\em J. Amer. Statist. Assoc.}, 109:411--423, 2014.

\bibitem[DMN92]{DMN92} 
M.B.~Dillencourt, D.M.~Mount, and N.S.~Netanyahu. %
\newblock A randomized algorithm for slope selection. 
\newblock {\em Internat. J. Comput. Geom. Appl.}, 2:1--27, 1992.

\bibitem[DG92]{DG92} 
D.L.~Donoho and M.~Gasko.%
\newblock Breakdown properties of location estimates based 
on halfspace depth and projected outlyingness.
\newblock {\em Ann. Statist.}, 20:1803--1827, 1992. 

\bibitem[DM14]{DM14}
R.~Dyckerhoff and P.~Mozharovskyi.
\newblock Exact computation of the halfspace depth.
\newblock arXiv:1411.6927 [stat.CO], 2014.

\bibitem[ES90]{ES90} 
H.~Edelsbrunner and D.L.~Souvaine.%
\newblock Computing least median of squares regression 
lines and guided topological sweep.
\newblock {\em J. Amer. Statist. Assoc.}, 85:115--119, 1990. 

\bibitem[Epp97]{Epp97} 
D.~Eppstein.
\newblock Faster construction of planar two-centers.
\newblock In {\em Proc. 8th Annu. ACM-SIAM Sympos. Discrete
          Algorithms}, pages 131--138, New Orleans, 1997.

\bibitem[Epp98]{Epp98} 
D.~Eppstein.
\newblock Fast hierarchical clustering and other 
applications of dynamic closest pairs.
\newblock In {\em Proc. 9th Annu. ACM-SIAM Sympos. Discrete
          Algorithms}, pages 619--628, San Francisco, 1998.

\bibitem[FFC12]{FFC12}
H.~Fritz, P.~Filzmoser, and C.~Croux.
\newblock A comparison of algorithms for the multivariate
          $L_1$-median.
\newblock {\em Computation. Stat.}, 27:393--410, 2012.

\bibitem[Gow74]{Gow74} 
J.C.~Gower.%
\newblock The mediancenter.
\newblock {\em J. Roy. Statist. Soc. Ser. C}, 32:466--470, 1974.

\bibitem[HPS10]{HPS10}
M.~Hallin, D.~Paindaveine, and M.~{\v{S}}iman.
\newblock Multivariate quantiles and multiple-output regression
          quantiles: from $L_1$-optimization to halfspace depth.
\newblock {\em Ann. Statist.}, 38:635--669, 2010.

\bibitem[HRV01]{HRV01}
M.~Hubert, P.J.~Rousseeuw, and S.~Van Aelst. %
\newblock Similarities between location depth and regression depth. 
\newblock In L.T.~Fernholz, editor, 
{\em Statistics in Genetics and in the Environmental Sciences}, 
pages 153--162.
\newblock Birkh\"{a}user Verlag, Basel, 2001. 

\bibitem[HRS15a]{HRS15a}
M.~Hubert, P.J.~Rousseeuw, and P.~Segaert.
\newblock Multivariate functional outlier detection. 
\newblock {\em Stat. Methods \& Applic.}, 24:177--202, 2015.

\bibitem[HRS15b]{HRS15b}
M.~Hubert, P.J.~Rousseeuw, and P.~Segaert.
\newblock Multivariate and functional classification using
          depth and distance.
\newblock arXiv:1504.01128, 2015. 

\bibitem[JM94]{JM94}
S.~Jadhav and A.~Mukhopadhyay. %
\newblock Computing a centerpoint of a finite planar set of points
in linear time.
\newblock {\em Discrete Comput. Geom.}, 12:291--312, 1994. 

\bibitem[KM12]{KM12}
L.~Kong and I.~Mizera.
\newblock Quantile tomography: using quantiles with
          multivariate data.
\newblock {\em Statistica Sinica}, 22:1589--1610, 2012.

\bibitem[KMV02]{KMV02} 
S.~Krishnan, N.H.~Mustafa, and S.~Venkatasubramanian.%
\newblock Hardware-assisted computation of depth contours.
\newblock In {\em Proc. 13th Annu. ACM-SIAM Sympos. Discrete
          Algorithms}, pages 558--567, 2002.

\bibitem[LS00]{LS00} 
S.~Langerman and W.~Steiger. %
\newblock Computing a maximal depth point in the plane. 
\newblock In {\em Proc. Japan Conf. on Discrete and 
          Computational Geometry}, pages 46--47, 2000.

\bibitem[LS00b]{LS00b} 
S.~Langerman and W.~Steiger. 
\newblock An optimal algorithm for hyperplane depth 
in the plane. 
\newblock In {\em Proc. 11th Annu. ACM-SIAM Sympos. 
Discrete Algorithms}, pages 54--59, San Francisco, 2000.

\bibitem[LS03]{LS03} 
S.~Langerman and W.~Steiger. 
\newblock Optimization in arrangements. 
\newblock {\em Lecture Notes in Computer Science 2607,
Springer-Verlag, Berlin, 2003.}

\bibitem[LS03b]{LS03b} 
S.~Langerman and W.~Steiger. 
\newblock The complexity of hyperplane depth in the plane. 
\newblock {\em Discrete Comput. Geom.}, 30:299--309, 2003. 

\bibitem[LCL12]{LCL12}
J.~Li, J.A.~Cuesta-Albertos, and R.Y.~Liu.
\newblock DD-classifier: nonparametric classification
          procedure based on DD-plot.
\newblock {\em J. Amer. Statist. Assoc.}, 107:737--753, 2012.				

\bibitem[Liu90]{Liu90} 
R.Y.~Liu. %
\newblock On a notion of data depth based on random simplices. 
\newblock {\em Ann. Statist.}, 18:405--414, 1990.

\bibitem[LPS99]{LPS99} 
R.Y.~Liu, J.~Parelius, and K.~Singh. %
\newblock Multivariate analysis by data depth: descriptive
          statistics, graphics and inference. 
\newblock {\em Ann. Statist.}, 27:783--840, 1999.

\bibitem[LMM14]{LMM14}
X.~Liu, K.~Mosler, and P.~Mozharovskyi.
\newblock Fast computation of Tukey trimmed regions in 
          dimension $p>2$.
\newblock arXiv:1412.5122 [stat.CO], 2014.

\bibitem[LZ14]{LZ14}
X.~Liu and Y.~Zuo.
\newblock Computing projection depth and its associated 
          estimators.
\newblock {\em Stat. Comput.}, 24:51--63, 2014.

\bibitem[MMN98]{MMN98}
J.~Matou\v{s}ek, D.M.~Mount, and N.S.~Netanyahu.
\newblock Efficient randomized algorithms for the repeated 
median line estimator.
\newblock {\em Algorithmica}, 20:136--150, 1998.

\bibitem[MRR{\etalchar{+}}03]{MRRp03}
K.~Miller, S.~Ramaswami, P.~Rousseeuw, J.A.~Sellar\`es, 
D.~Souvaine, I.~Streinu, and A.~Struyf.%
\newblock Efficient Computation of Location Depth Contours
by Methods of Computational Geometry.
\newblock {\em Stat. Comput.}, 13:153--162, 2003.

\bibitem[Miz02]{Miz02}
I.~Mizera.
\newblock On depth and deep points: a calculus.
\newblock {\em Ann. Statist.} 30:1681--1736, 2002.

\bibitem[Mos13]{Mos13}
K.~Mosler.
\newblock Depth statistics.
\newblock In C. Becker, R. Fried, and S. Kuhnt, editors,
          {\em Robustness and Complex Data Structures},
					pages 17--34,
\newblock Springer-Verlag, Berlin Heidelberg, 2013.

\bibitem[MN94]{MN94}
D.M.~Mount and N.S.~Netanyahu.
\newblock Computationally efficient algorithms for 
          high-dimensional robust estimators.
\newblock {\em Graphical Models Image Proc.}, 56:289--303, 1994.

\bibitem[MN{\etalchar{+}}97]{MNRp97}
D.M.~Mount, N.S.~Netanyahu, K.~Romanik, R.~Silverman, and A.Y.~Wu.%
\newblock A practical approximation algorithm for the LMS line estimator.
\newblock In {\em Proc. 8th Annu. ACM-SIAM Sympos. Discrete 
          Algorithms}, pages 473--482, New Orleans, 1997.

\bibitem[Oja83]{Oja83}
H.~Oja.%
\newblock Descriptive statistics for multivariate distributions.
\newblock {\em Statist. Probab. Lett.}, 1:327--332, 1983.

\bibitem[PK97]{PK97}
S.~Portnoy and R.~Koenker.
\newblock The Gaussian hare and the Laplacian tortoise: 
  computability of squared-error
	versus absolute-error estimators.
\newblock {\em Stat. Sci.,} 12:279--300, 1997.

\bibitem[Rou84]{Rou84} 
P.J.~Rousseeuw. 
\newblock Least median of squares regression. 
\newblock {\em J. Amer. Statist. Assoc.}, 79:871--880, 1984.

\bibitem[RH99]{RH99} 
P.J.~Rousseeuw and M.~Hubert. 
\newblock Regression depth.
\newblock {\em J. Amer. Statist. Assoc.}, 94:388--402, 1999.

\bibitem[RH99b]{RH99b}
P.J.~Rousseeuw and M.~Hubert. 
\newblock Depth in an arrangement of hyperplanes.
\newblock {\em Discrete Comput. Geom.}, 22:167--176, 1999. 

\bibitem[RR96]{RR96}
P.J.~Rousseeuw and I.~Ruts.%
\newblock Algorithm AS 307: Bivariate location depth.
\newblock {\em J. Roy. Statist. Soc. Ser. C}, 45:516--526, 1996.

\bibitem[RR98]{RR98}
P.J.~Rousseeuw and I.~Ruts.
\newblock Constructing the bivariate Tukey median.
\newblock {\em Statistica Sinica}, 8:827-839, 1998.

\bibitem[RRT99]{RRT99}
P.J.~Rousseeuw, I.~Ruts, and J.W.~Tukey.%
\newblock The bagplot: A bivariate boxplot.
\newblock {\em Amer. Statist.}, 53:382--387, 1999. 

\bibitem[RS98]{RS98} 
P.J.~Rousseeuw and A.~Struyf. 
\newblock Computing location depth and regression depth 
in higher dimensions.
\newblock {\em Stat. Comput.}, 8:193--203, 1998.

\bibitem[RS04]{RS04}
P.J.~Rousseeuw and A.~Struyf. 
\newblock Characterizing angular symmetry and regression symmetry.
\newblock {\em J. Statist. Plann. Inference}, 122:161-–173, 2004.
 
\bibitem[RR96b]{RR96b}
I.~Ruts and P.J.~Rousseeuw.
\newblock Computing depth contours of bivariate point clouds.
\newblock {\em Comput. Stat. Data Anal.}, 23:153--168, 1996.

\bibitem[Sen68]{Sen68} 
P.K.~Sen. %
\newblock Estimates of the regression coefficient based on 
          Kendall's tau.
\newblock {\em J. Amer. Statist. Assoc.}, 63:1379--1389, 1968.

\bibitem[Ser02]{Ser02}
R.~Serfling.
\newblock A depth function and a scale curve based on spatial quantiles.
\newblock In Y.~Dodge, editor, {\em Statistical Data Analysis Based on the L1-Norm and Related Methods},
pages 25--38,
\newblock Birkha\"{u}ser, Basel, 2002.

\bibitem[Sha76]{Sha76}
M.I.~Shamos.%
\newblock Geometry and statistics: problems at the interface.
\newblock In J.F.~Traub, editor, {\em Algorithms and 
          Complexity: New Directions and Recent Results}, 
          pages 251--280.
\newblock Academic Press, Boston, 1976. 

\bibitem[Sie82]{Sie82} 
A.~Siegel. %
\newblock Robust regression using repeated medians.
\newblock {\em Biometrika}, 69:242--244, 1982.

\bibitem[Sma90]{Sma90} 
C.G.~Small. %
\newblock A survey of multidimensional medians.
\newblock {\em Internat. Statistical Review}, 58:263--277, 1990.

\bibitem[SHR97]{SHR97}
A.~Struyf, M.~Hubert, and P.J.~Rousseeuw.
\newblock Integrating robust clustering techniques in S-PLUS.
\newblock {\em Comput. Stat. Data Anal.}, 26:17--37, 1997.

\bibitem[SR00]{SR00}
A.~Struyf and P.J.~Rousseeuw.
\newblock High-dimensional computation of the deepest location.
\newblock {\em Comput. Stat. Data Anal.}, 34:415--426, 2000.

\bibitem[The50]{The50}
H.~Theil.%
\newblock A rank-invariant method of linear and polynomial 
regression analysis (parts 1-3).
\newblock {\em Nederl. Akad. Wetensch. Ser. A}, 53:386--392,
          521--525, 1397--1412, 1950.

\bibitem[Tuk75]{Tuk75}  
J.W.~Tukey. 
\newblock Mathematics and the picturing of data.
\newblock In {\em Proc. Internat. Congr. of Math.}, 2, 
          pages 523--531, Vancouver, 1975.

\bibitem[VMR{\etalchar{+}}08]{VMRp08}
M.~van Kreveld, J.~Mitchell, P.~Rousseeuw, M.~Sharir,
J.~Snoeyink, B.~Speckmann.
\newblock Efficient algorithms for maximum regression depth.
\newblock {\em Discrete Comput. Geom.}, 39:656--677, 2008.

\bibitem[VRH{\etalchar{+}}02]{VRHp02}
S.~Van Aelst, P.J.~Rousseeuw, M.~Hubert, and A.~Struyf.
\newblock The deepest regression method.
\newblock {\em J. Multivar. Anal.}, 81:138--166, 2002.

\bibitem[YKI{\etalchar{+}}88]{YKIp88}
P.~Yamamoto, K.~Kato, K.~Imai, and H.~Imai.%
\newblock Algorithms for vertical and orthogonal $L^1$ linear 
approximation of points. 
\newblock In {\em Proc. 4th Sympos. Comput. Geom.}, 
          pages 352--361, 1988.

\bibitem[Zuo03]{Zuo03} 
Y.~Zuo.%
\newblock Projection based depth functions and associated medians.
\newblock {\em Ann. Statist.}, 31:1460--1490, 2003.

\bibitem[ZS00]{ZS00} 
Y.~Zuo and R.~Serfling.%
\newblock General notions of statistical depth function. 
\newblock {\em Ann. Statist.}, 28:461--482, 2000.

\end{thebibliography}
%
\newcommand{\etalchar}[1]{$^{#1}$}

\end{document}